\documentclass[aps,prl,twocolumn,reprint,showpacs,superscriptaddress]{revtex4-1}
\usepackage{graphicx}
\usepackage{dcolumn}
\usepackage[lmargin=1.5cm, rmargin=1.5cm,tmargin=1.5cm,bmargin=2.0cm]{geometry}
\usepackage{bm}
\usepackage{textcomp}
\usepackage{xcolor}
\usepackage{ulem}
\usepackage{booktabs}
\usepackage{longtable}
\usepackage{multirow}
\usepackage{amsmath} 
\usepackage{amssymb}
\RequirePackage{hyperref}
\hypersetup
{
  bookmarks=true,
  unicode=false,
  pdftoolbar=true,
  pdfmenubar=true,
  pdffitwindow=false,
  pdfstartview=FitH,
  pdftitle={PRC},
  pdfauthor={SU Jun},
  pdfsubject={Nuclear Physics},
  pdfcreator={SU Jun},
  pdfproducer={SU Jun},
  pdfkeywords={IMME},		
  bookmarksnumbered,
  colorlinks=true,
  linkcolor=blue,
  citecolor=red,
  filecolor=green
  urlcolor=blue,
  breaklinks,
  plainpages=false
}

\usepackage{lineno}
 
\begin{document}


\preprint{APS/123-QED}

\title{Evidence for enhanced neutron-proton correlations from the level structure of the $N=Z+1$ nucleus $^{87}_{43}$Tc$^{\ }_{44}$}

\newcommand{\kth} {\affiliation{Department of Physics, Royal Institute of Technology, Stockholm 104 05, Sweden}}
\newcommand{\imp} {\affiliation{Institute of Modern Physics, Chinese Academy of Sciences, Lanzhou 730000, China}}
\newcommand{\ucas}{\affiliation{University of Chinese Academy of Sciences, Beijing 100049, China}}
\newcommand{\Istanbul} {\affiliation{Department of Physics, Faculty of Science, Istanbul University, Vezneciler/Fatih, 34134 Istanbul, Turkey}}
\newcommand{\GANIL} {\affiliation{GANIL, CEA/DRF-CNRS/IN2P3, Boulevard Henri Becquerel, BP 55027, F-14076 Caen Cedex 5, France}}
\newcommand{\CNRS} {\affiliation{Centre de Sciences Nucl\`eaires et Sciences de la Mati\`ere, CNRS/IN2P3, Universit\`e Paris-Saclay, 91405 Orsay, France}}
\newcommand{\CNRSI} {\affiliation{Universit\'e Paris-Saclay, CNRS/IN2P3, IJCLab, 91405 Orsay, France}}
\newcommand{\CSICb} {\affiliation{Instituto de F\`isica Corpuscular, CSIC-Universidad de Valencia, E-46980 Valencia, Spain}}
\newcommand{\Nazionali} {\affiliation{Istituto Nazionale di Fisica Nucleare, Laboratori Nazionali di Legnaro, I-35020 Legnaro, Italy}}
\newcommand{\Warsaw} {\affiliation{Heavy Ion Laboratory, University of Warsaw, ul. Pasteura 5A,02-093 Warszawa, Poland}}
\newcommand{\Atomki} {\affiliation{Institute for Nuclear Research, Atomki, H-4001 Debrecen, Hungary}}
\newcommand{\Uppsala} {\affiliation{Department of Physics and Astronomy, Uppsala University, SE-75121 Uppsala, Sweden}}
\newcommand{\York} {\affiliation{Department of Physics, University of York, Heslington, York, YO10 5DD, United Kingdom}}
\newcommand{\Rustaq} {\affiliation{Rustaq College of Education, Department of Science, 329 Al-Rustaq, Sultanate of Oman}}
\newcommand{\Oliverb} {\affiliation{Department of Physics, Oliver Lodge Laboratory, University of Liverpool, Liverpool L69 7ZE, United Kingdom}}
\newcommand{\Lyonb} {\affiliation{Universit\`e Lyon, CNRS/IN2P3, IPN-Lyon, F-69622, Villeurbanne, France}}
\newcommand{\CERN} {\affiliation{CERN, CH-1211 Geneva 23, Switzerland}}
\newcommand{\Henryk} {\affiliation{The Henryk Niewodniczański Institute of Nuclear Physics, Polish Academy of Sciences,ul. Radzikowskiego 152, 31-342 Kraków, Poland}}
\newcommand{\INFNs} {\affiliation{INFN Sezione di Milano, I-20133 Milano, Italy}}
\newcommand{\Kernphysik} {\affiliation{Institut f\"ur Kernphysik, Universit\"at zu K\"oln, Z\"ulpicher Str. 77, D-50937 K\"oln, Germany}}
\newcommand{\Olivera} {\affiliation{Oliver Lodge Laboratory, The University of Liverpool, Liverpool, L69 7ZE, United Kingdom}}
\newcommand{\STFC} {\affiliation{STFC Daresbury Laboratory, Daresbury, Warrington, WCNRS 4AD, United Kingdom}}
\newcommand{\IPHC} {\affiliation{IPHC, UNISTRA, CNRS, 23 rue du Loess, 67200 Strasbourg, France}}
\newcommand{\Milano} {\affiliation{University of Milano, Department of Physics, I-20133 Milano, Italy}}
\newcommand{\INFNm} {\affiliation{INFN Milano, I-20133 Milano, Italy}}
\newcommand{\Manchester} {\affiliation{Nuclear Physics Group, Schuster Laboratory, University of Manchester, Manchester, M13 9PL, United Kingdom}}
\newcommand{\Campus} {\affiliation{CNRS-IN2P3, Universite\`e Paris-Saclay, Bat 104, F-91405 Orsay Campus, France}}
\newcommand{\CSICv} {\affiliation{Instituto de F\`isica Corpuscular, CSIC-Universidad de Valencia, E-46071 Valencia, Spain}}
\newcommand{\Zaim} {\affiliation{Faculty of Engineering and Natural Sciences, Istanbul Sabahattin Zaim University, 34303, Istanbul, Turkey}}
\newcommand{\Nigde} {\affiliation{Department of Physics, University of Nigde, 51240 Nigde, Turkey}}
\newcommand{\Valencia} {\affiliation{Departamento de Ingenier\`ia Electr\`onica, Universitat de Valencia, 46100 Burjassot, Valencia, Spain}}
\newcommand{\CSICm} {\affiliation{Instituto de Estructura de la Materia, CSIC, Madrid, E-28006 Madrid, Spain}}
\newcommand{\CEA} {\affiliation{Irfu, CEA, Universi\`e Paris-Saclay, F-91191 Gif-sur-Yvette, France}}
\newcommand{\INFNPa} {\affiliation{INFN Padova, I-35131 Padova, Italy}}
\newcommand{\Surrey} {\affiliation{Department of Physics, University of Surrey, Guildford, GU2 7XH, United Kingdom}}
\newcommand{\INFNp} {\affiliation{Dipartimento di Fisica e Astronomia dell’Universit\`a di Padova and INFN Padova, I-35131 Padova, Italy}}
\newcommand{\Surreya} {\affiliation{Department of Physics, University of Surrey, Guildford, GU2 7XH, United Kingdom}}
\newcommand{\bLyon} {\affiliation{Dipartimento di Fisica e Astronomia dell’Universit\`a di Padova and INFN Padova, I-35131 Padova, Italy}}
\newcommand{\Lyon} {\affiliation{Universit\`e Lyon 1, CNRS/IN2P3, IPN-Lyon, F-69622, Villeurbanne, France}}
\newcommand{\SU} {\affiliation{Department of Physics, Stockholm University, Stockholm 106 91, Sweden}}

\author{X.~Liu} \email{xiaoyuli@kth.se} \kth \imp \ucas
\author{B. Cederwall} \kth
\author{C. Qi} \kth
\author{R. A. Wyss} \kth
\author{\"O. Aktas} \kth
\author{A. Ertoprak} \kth \Istanbul
\author{W. Zhang} \kth
\author{E. Cl\'ement} \GANIL
\author{G. de France} \GANIL
\author{D. Ralet} \CNRS
\author{A. Gadea} \CSICb
\author{A. Goasduff} \Nazionali
\author{G. Jaworski} \Nazionali \Warsaw
\author{I. Kuti} \Atomki
\author{B. M. Nyak\'o} \Atomki
\author{J. Nyberg} \Uppsala
\author{M. Palacz} \Warsaw
\author{R. Wadsworth} \York
\author{J. J. Valiente-Dob\'on} \Nazionali
\author{H. Al-Azri} \Rustaq
\author{A. Ata\ifmmode \mbox{\c{c}}\else \c{c}\fi{} Nyberg} \kth
\author{T. B\"ack} \kth
\author{G. de Angelis} \Nazionali
\author{M. Doncel} \Olivera \SU
\author{J. Dudouet} \Lyonb
\author{A. Gottardo} \CNRS
\author{M. Jurado} \CSICb
\author{J. Ljungvall} \CNRS
\author{D. Mengoni} \Nazionali
\author{D. R. Napoli} \Nazionali
\author{C. M. Petrache} \CNRSI
\author{D. Sohler} \Atomki
\author{J. Tim\'ar} \Atomki
\author{D. Barrientos} \CERN
\author{P. Bednarczyk} \Henryk
\author{G. Benzoni} \INFNm 
\author{B. Birkenbach} \Kernphysik
\author{A. J. Boston} \Olivera
\author{H. C. Boston} \Olivera
\author{I. Burrows} \STFC
\author{L. Charles} \IPHC
\author{M. Ciemala} \Henryk
\author{F. C. L. Crespi} \Milano \INFNm
\author{D. M. Cullen} \Manchester
\author{P. D\'esesquelles} \Campus \CNRS
\author{C. Domingo-Pardo} \CSICb 
\author{J. Eberth} \Kernphysik
\author{N. Erduran} \Zaim
\author{S. Ert\"urk} \Nigde
\author{V. Gonz\'alez} \Valencia
\author{J. Goupil} \GANIL
\author{H. Hess} \Kernphysik
\author{T. Huyuk} \CSICb
\author{A. Jungclaus} \CSICm
\author{W. Korten} \CEA
\author{A. Lemasson} \GANIL
\author{S. Leoni} \Milano \INFNm
\author{A. Maj} \Henryk
\author{R. Menegazzo} \INFNPa
\author{B. Million} \INFNm
\author{R. M. Perez-Vidal} \CSICb 
\author{Zs. Podoly\`ak} \Surrey
\author{A. Pullia} \Milano \INFNm
\author{F. Recchia} \INFNPa 
\author{P. Reiter} \Kernphysik
\author{F. Saillant} \GANIL
\author{M. D. Salsac} \CEA
\author{E. Sanchis} \Valencia
\author{J. Simpson} \STFC
\author{O. Stezowski} \Lyonb 
\author{C. Theisen} \CEA
\author{M. Zieli\ifmmode \acute{n}\else \'{n}\fi{}ska} \CEA

\date{\today}

\begin{abstract}

The low-lying excited states in the neutron-deficient $N=Z+1$ nucleus $^{87}_{43}$Tc$^{\ }_{44}$ have been studied via the fusion-evaporation reaction $^{54}$Fe($^{36}$Ar, $2n1p$)$^{87}$Tc at the Grand Acc\'el\'erateur National d'Ions Lourds (GANIL), France. The AGATA spectrometer was used in conjunction with the auxiliary NEDA, Neutron Wall, and DIAMANT detector arrays to measure coincident prompt $\gamma$-rays, neutrons, and charged particles emitted in the reaction. A level scheme of $^{87}$Tc from the (9/2$^{+}_{g.s.}$) state to the (33/2$^{+}_{1}$) state was established based on 6 mutually coincident $\gamma$-ray transitions. The constructed level structure exhibits a rotational behavior with a sharp backbending at $\hbar\omega\approx 0.50$ MeV. A decrease in alignment frequency and increase in alignment sharpness in the odd-mass isotonic chains around $N=44$ is proposed as an effect of the enhanced isoscalar neutron-proton interactions in odd-mass nuclei when approaching the $N=Z$ line.

\end{abstract}


\maketitle

\paragraph{Introduction.}

The nuclear pairing correlation is an essential ingredient for describing the properties of finite atomic nuclei. The underlying mechanism is considered to be analogous to that formulated in the Bardeen-Cooper-Schrieffer (BCS) theory~\cite{bard1957,bohr1958}. Unique for the nuclear system is that it is a strongly correlated system of two different types of fermions (neutrons and protons). Due to isospin symmetry, the Cooper pairs can be formed in both the $T=1$ isovector (neutron-neutron, proton-proton, and neutron-proton) channel and the $T=0$ isoscalar (neutron-proton) channel. A wealth of experimental observations have supported the fundamental role of neutron-neutron and proton-proton pairing correlations in understanding various nuclear properties including mass, deformation, moment of inertia and rotational alignment~\cite{brog2013, dean2003}. The effect of $T=1$ isovector neutron-proton ($np$) pairing has also been analyzed from different perspectives~\cite{macc2000, good1979, civi1997, enge1997, glow2004}. An interesting problem that is still considered open is the experimental evidence for the  $T=0$ isoscalar $np$ pairing mode. The possible existence of such deuteron-like pairing has attracted extensive studies over the years~\cite{civi1997, enge1997, frau2014, satu1997}. 

It is expected that the effects of isoscalar pairing can be enhanced along the $N=Z$ line where the valence neutrons and protons occupy identical orbitals, in particular in the heaviest self-conjugate systems where many particles might contribute to form an isoscalar pairing condensate. In recent years, advances in detection and data acquisition technologies have opened up the nuclei far from stability to observation, and the newly unveiled spectroscopy data in the intermediate-mass $N\sim Z$ region has reignited the interest in investigating the possible manifestation of the $T=0$ pairing mode. The occurrence of a significant component of isoscalar spin-aligned $np$ pairs in the nuclear wave function is suggested in the heavy spherical $N=Z$ nuclei like $^{92}$Pd~\cite{cede2011, satu1997}. The effect of the competition between the spin-aligned $np$ pair interaction and the quadrupole correlation could be strong even in the lighter, modestly deformed, nucleus $^{88}$Ru~\cite{kane2017, cede2020}. It is the response of the different pairing field components to collective rotation~\cite{satu1997, tera1998, shei2000} that is of interest in such deformed systems. A possible consequence of the increase in rotational frequency is that while the $T=1$ isovector pairs, which couple two nucleons to $J=0$, are successively destroyed by the increasing Coriolis force, the $T=0$ isoscalar pairs with $J>0$ can align to build up angular momentum~\cite{frau2014} and thus may still leave fingerprints after the quenching of $T=1$ pairing. The ground-state rotational bands of even-even $N=Z$ nuclei from $^{72}$Kr to $^{88}$Ru~\cite{fisc2001, marg2002, marg2001, cede2020} have been extended to the region where band crossings are normally expected, and increases in crossing frequency compared to their neighboring $N>Z$ nuclei were consistently observed. These ``delayed" alignments have been widely suggested as arising from the $T=0$ $np$ pairing correlation~\cite{ange1997}, though the effect of the shape degrees of freedom can muddle the analysis in some cases. 

The effect of $T=0$ $np$ pairing correlations may be strong also for $N=Z+1$ nuclei even though the impact is expected to diminish rapidly as one goes away from the $N=Z$ line~\cite{cede2020}. In the rotational odd-mass nuclei, the first rotational alignment can be explicitly assigned to either aligned neutrons or aligned protons due to the odd-particle blocking in the isovector pairing field. After the crossing to the 3-quasiparticle band, due to the simultaneous presence of unpaired neutrons and protons at high spin will be favored by the $T=0$ $np$ pairing field. In the $T_{z}=\pm 1/2$ nuclei, such a competition may lead to a transition between the two pairing phases. In Ref. \cite{wyss2007}, it is argued that the negative parity bands in $^{73}$Kr can be accounted for by means of $T = 0$ pair correlations. The deformed region around $^{88}$Ru is an ideal laboratory to verify the condensation of isoscalar pairs due to the fact that most valence particles are confined in the high angular momentum \textsl{g}$_{9/2}$ orbital below the magic number $N=Z=50$. The intermediate-mass $T_{z}=1/2$ nuclei have been mapped up to $^{95}$Ag~\cite{marg2003} with relatively rich spectroscopic information. The exception is $^{87}$Tc, for which only two excited states were previously known~\cite{rudo1991}. In this study, we report on the observation of excited states in the yrast band of $^{87}$Tc from the (9/2$^{+}_{g.s.}$) state to the (33/2$^{+}_{1}$) state. The results are discussed based on the systematical studies and compared with large-scale and single-$j$ shell model calculations including the possible effects from the $T=0$ $np$ pairing correlation.

\paragraph{Experiment.}

The experiment was performed at the Grand Acc\'el\'erateur National d'Ions Lourds (GANIL), Caen, France. The reaction $^{54}$Fe($^{36}$Ar, $2n1p$)$^{87}$Tc was induced by a 5 $\sim$ 10 pnA $^{36}$Ar beam at 115 MeV which was led to bombard the 6 mg/cm$^{2}$ thick $^{54}$Fe target foils. Prompt $\gamma$ rays from the de-excitation of the rare $^{87}$Tc nuclei produced in the reaction were measured with the Advanced Gamma Tracking Array (AGATA)~\cite{clem2017} spectrometer consisting of 11 triple-cluster segmented HPGe detectors~\cite{akko2012}. The indispensable channel selection was achieved by operating AGATA in conjunction with auxiliary detector systems for light particles. Evaporated charged particles were recorded in the DIAMANT~\cite{sche1997, jgal2004} detector array consisting of 60 CsI(Tl) scintillators placed inside the target chamber. In the forward hemisphere, with approximately 1.6$\pi$ solid angle coverage, the NEDA~\cite{Huyu2016, jvdo2019} and Neutron Wall~\cite{skep1999} detector arrays, consisting of 54 and 42 organic liquid scintillator detectors, respectively, were placed to detect the emitted neutrons. AGATA was calibrated with a standard $^{152}$Eu radioactive source giving a 0.095$\%$ $\sigma/E$ resolution for the 1408.1 keV $\gamma$-ray transition. The trigger condition in the experiment required at least 1 (NEDA or Neutron Wall) neutron and 2 (AGATA) $\gamma$-ray detectors to fire in coincidence.

\begin{figure}[htpb]
\includegraphics[width=0.485\textwidth]{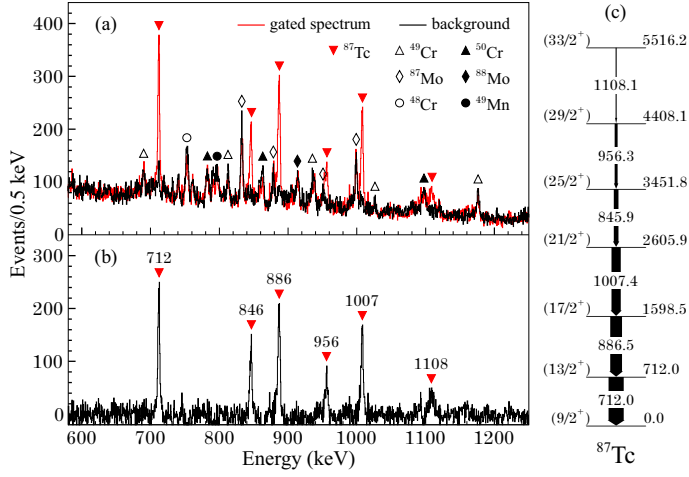}
\caption{(a) Sum of the spectra gated on the 712.0, 886.5, 1007.4, 845.9 and 956.3 keV lines (red), and the summed background spectrum (black) gated 4 keV above these lines. The background spectrum was normalized to the gated spectrum according to the statistics below 2 MeV. $\gamma$-ray peaks due to contaminant reactions are indicated. The coincidence spectrum in (b) is the difference between the two spectra shown in (a). (c) Level scheme of $^{87}$Tc deduced from the present work. The tentative spin-parity assignments were made based on the systematics as well as shell-model calculations, and the width of the arrows are proportional to the relative intensities of the $\gamma$ rays. The relative intensities: 100(4), 80(3), 65(5), 37(4), 26(3), and 13(3).}
\label{spectra}
\end{figure}

In the off-line analysis, $\gamma$-ray events belonging to the $2n1p$ channel were identified by using the information on the detected light particles. The $\gamma$-$\gamma$ coincidence matrix for $^{87}$Tc was sorted with the condition of $``2n01p"$, which means that two neutrons together with either none or one proton had to be detected in prompt coincidence. The acceptance of $``0p"$ events is mainly due to the low single-proton detection efficiency of only about 39(1)\%. In order to suppress the contamination from $1n$ channels, a purification of the $``2n"$ events was performed by rejecting the neutron-scattering events using the time-of-flight of the detected neutrons and the positions of the detectors that fired. The requirement of $``01p"$ events was either a complete silence of the whole DIAMANT array or the only signal must be assigned to a proton. In the $\gamma$-$\gamma$ coincidence matrix, a mutually coincident cascade of six transitions was observed (see Fig.~\ref{spectra}). In addition to the first two transitions at 712.0(1) and 886.5(1) keV which have been reported previously~\cite{rudo1991}, four new lines at 1007.4(1), 845.9(1), 956.3(2), and 1108.1(4) keV were identified. The background-subtracted spectrum is shown in Fig.~\ref{spectra} (b) in which all six peaks are clearly visible. A level scheme of $^{87}$Tc, established based on the coincidence relationships and the relative intensities of the $\gamma$ rays, is displayed in Fig.~\ref{spectra} (c). 

\paragraph{Discussion.}

\begin{figure}
\includegraphics[width=0.44\textwidth]{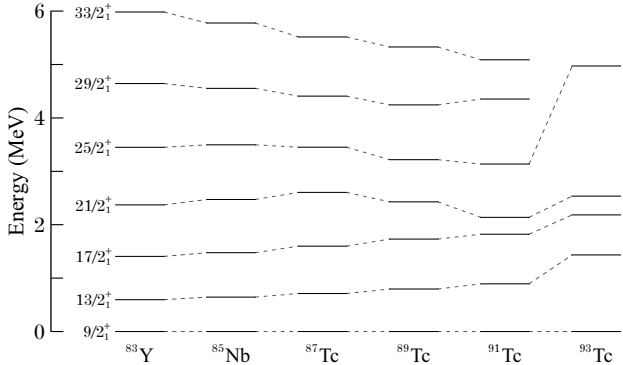}
\caption{Systematics of positive-parity yrast states below 6 MeV in $^{87}$Tc together with the odd-A $N=44$ isotones~\cite{cris1992, gros1991} in comparison with neighboring technetium isotopes $^{89,91,93}$Tc~\cite{rudo1995, ghug1993, rudo1994}. The results for $^{87}$Tc are from the present work.}
\label{isotopn}
\end{figure}

The systematics of positive-parity yrast states below 6 MeV in the odd-$A$ $N=44$ isotones and those in the neighboring technetium isotopes $^{89,91,93}$Tc are shown in Fig.~\ref{isotopn}. The $N=44$ isotones show a characteristic rotational behavior while a strong transition to spherical coupling is observed along the Tc isotopic chain as one approaches $N=50$. In the semi-magic nucleus $^{93}$Tc, the level pattern up to 21/2$_{1}^{+}$ is of a typical spherical $\pi$(\textsl{g}$_{9/2}$)$^{3}$ multiplet character. Here, a sudden and significant increase in the excitation energy is seen for the 25/2$_{1}^{+}$ state which involves the excitation and breaking of a proton pair and has the configuration $\pi$(\textsl{g}$_{9/2}$)$^{5}_{25/2}$. As expected, the energy gaps between the 25/2$_{1}^{+}$ and 21/2$_{1}^{+}$ states are reduced in $^{89,91}$Tc in the presence of both valence protons and neutrons. In the middle of the $\nu$(\textsl{g}$_{9/2}$) subshell, the level pattern of $^{87}$Tc is characterized by a fully developed rotational band similar to the other $N=44$ isotones.

\begin{figure}
\includegraphics[width=0.485\textwidth]{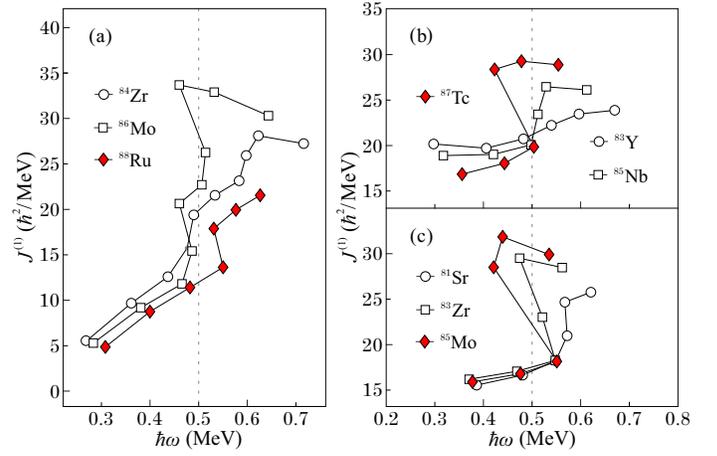}
\caption{Kinematic moments of inertia ($J^{(1)}$) as a function of rotational frequency for the positive parity and positive signature bands in the $N=44$ even-even~\cite{pric1983, rudo1996, cede2020} isotones (a), $N=44$ odd-mass isotones (b), and $N=43$ odd-mass~\cite{mar12002, smal1990, john1994} isotones (c).}
\label{alignment}
\end{figure}

Focusing on the rotational patterns of the $N=44$ isotones with $T_{z}=5/2$ ($^{83}$Y), 3/2 ($^{85}$Nb), and 1/2 ($^{87}$Tc), a gradual decrease in the collectivity of the 1-quasiparticle (1-qp) bands from $^{83}$Y to $^{87}$Tc may be expected as can be inferred from the increasing gap between the $9/2^{+}_{1}$ and $13/2^{+}_{1}$ states. On the other hand, the higher-spin states starting from $J=21/2$ clearly show an opposite trend. In order to clarify this striking feature we have extracted the kinematic moments of inertia, $J^{(1)}$~\cite{john1971}, as a function of rotational frequency for the $N=44$ isotones from $^{83}$Y to $^{88}$Ru, as illustrated in Fig.~\ref{alignment} (a) and (b). The even-even $N=44$ isotones $^{84}$Zr~\cite{moun1992} and $^{86}$Mo~\cite{rudo1996} show a \textsl{g}$_{9/2}$ proton alignment at $\hbar\omega\approx 0.48$ MeV followed by a neutron crossing around 0.6 MeV and 0.5 MeV, respectively. All odd-$A$ isotones in the figure show $\nu\textsl{g}_{9/2}$ quasiparticle alignments~\cite{cris1992, gros1991} around $\hbar\omega=0.5$ MeV with a clear and unexpected increases in the crossing sharpness with the decrease of $T_{z}$. The first alignment in the even-even nuclei shows an upward trend in frequency from $Z=$ 42 to 44. The considerably ``delayed" alignment in self-conjugate $^{88}_{44}$Ru$^{\ }_{44}$ was suggested to be associated with isoscalar pairing correlations~\cite{cede2020, kane2017}. Considering now also the $N=43$ isotones with $T_{z}=5/2$ ($^{81}$Sr), 3/2 ($^{83}$Zr), and 1/2 ($^{85}$Mo) (see Fig.~\ref{alignment} (c)), all the odd-mass isotones reveal the same trend: the band crossings become earlier and sharper when approaching the $N=Z$ line regardless of the nature of these alignments.

\begin{figure}
\includegraphics[width=0.485\textwidth]{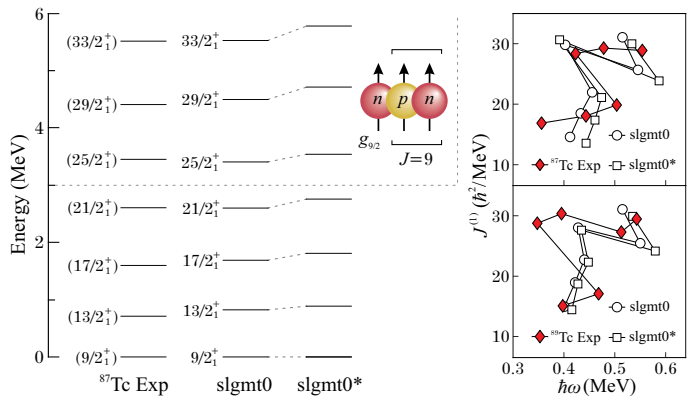}
\caption{Left: The experimental level scheme of $^{87}$Tc in comparison with the shell-model calculations in the $p\textsl{g}$ model space based on the slgmt0~\cite{serd1976} interaction (slgmt0* denotes the calculation with the spin-aligned neutron-proton interaction reduced by 200 keV). The cartoon illustrates the coupling between the unpaired neutron with the spin-aligned np pair which is the dominant component of the wave function for the higher-spin states. Right: Kinematic moments of inertia as a function of rotational frequency for the positive parity and positive signature bands in $^{87}$Tc and $^{89}$Tc from the experimental data and theoretical calculations (slgmt0 and slgmt0*).}
\label{level} 
\end{figure}

In Fig.~\ref{level} (left), we compare the observed level scheme of $^{87}$Tc with those from shell-model calculations in the $p\textsl{g}$ model space based on the effective interaction slgmt0~\cite{serd1976}. A calculation in the large-scale $fp\textsl{g}$ space with the interaction jun45~\cite{honm2009} was also performed and no significant difference between calculations in the two model spaces is expected since the wave functions for those states are dominated by the occupation of the $\textsl{g}_{9/2}$ orbitals. A slight modification of the lowest-lying states may be expected by the inclusion of the $d_{5/2}$ orbital as was shown in the case of the even-even $^{88}$Ru~\cite{kane2017}. For the higher-lying states from spin $J=25/2$, our calculations indicate that the coupling $\pi\textsl{g}_{9/2} \otimes (\pi\nu\textsl{g}_{9/2})_{9}$ is strongly favored in energy in relation to the fact that the spin-aligned $np$ pairs are more strongly attractive than other pairs, as discussed in \cite{cede2011,qiii2011, cede2020}. It leads to a fully aligned 3-qp-like structure at the 25/2$^{+}_{1}$ state as a result of the coupling between one unpaired neutron and a spin-aligned $np$ pair formed by the last odd proton and the other unpaired neutron. Higher-lying states can be built on top of that by the coupling to additional valence $np$ pairs. To illustrate this picture, as in Ref.~\cite{qiii2011}, we have redone our shell-model calculations in the $p\textsl{g}$ and $fp\textsl{g}$ model spaces by reducing the strength of the spin-aligned $np$ pair interaction. As an example, in the right part of Fig.~\ref{level} we have plotted calculations in the $p\textsl{g}$ space with the $(\textsl{g}_{9/2})^{2}_{J=9}$ spin-aligned $np$ pair interaction matrix element reduced slightly by 200 keV (denoted as slgmt0*). The reduced $np$ pair matrix element leads to an increase in the excitation energy for the higher-spin states much more significant than that of the low-lying states due to the enhanced $np$ pair coupling in their wave function. The increment is as much as 133 keV for the 25/2$^{+}_{1}$ state and nearly 200 keV for the states with higher spin. Our calculations furthermore show that the effect is much less pronounced when one goes away from $N=Z$. For example, in $^{89}$Tc, calculation with the same reduced $np$ pair matrix element leads only to a change of only 70 keV in the excitation energy of its 25/2$^{+}_{1}$ state. That is because the spin-aligned $np$ pair plays even larger role in $^{87}$Tc than in $^{89}$Tc. The experimental kinematic moments of inertia as a function of rotational frequency for $^{87}$Tc and $^{89}$Tc are compared with the shell-model calculations in Fig. 4 (right). The results of the calculations illustrate the importance of the spin-aligned neutron-proton coupling for the $T_{z}$ = 1/2 nucleus $^{87}$Tc as compared with $^{89}$Tc ($T_{z}$ = 3/2). In the latter case a reduction of the spin-aligned neutron-proton interaction energy by 200 keV has little influence on its structure.

Based on the results above, we suggest that the enhancement of the isoscalar $np$ correlations approaching the $N=Z$ line will make such spin-aligned states more energetically favored, leading to an increasing collectivity after the band crossing. Hence, in the presence of strong isoscalar $np$ correlation the aligned $np$ pair can result in a significant structural change between the 3-qp band and the 1-qp band within the conventional pairing scheme, which will additionally reduce the interaction strength between them and consequently induce a sharper alignment in line with the experimental observations. 

\paragraph{Conclusions.}

In summary, we have studied the low-lying yrast states in $^{87}$Tc via the $^{54}$Fe($^{36}$Ar, $2n1p$)$^{87}$Tc fusion-evaporation reaction at the GANIL accelerator complex. The prompt $\gamma\gamma$-neutron and charged-particle coincidences were measured by using the AGATA $\gamma$-ray spectrometer with the auxiliary NEDA, Neutron Wall, and DIAMANT arrays, resulting in an extension of the known $^{87}$Tc level scheme by 8 units of angular momentum to the tentative $33/2^{+}_{1}$ state with the first band crossing included. The observed level structure is compared with the neighboring odd-mass $N=44$ and $43$ isotonic chains as well as shell-model calculations. A striking feature of decreasing excitation energy in the higher spin states starting from 25/2$^{+}$ is observed when approaching the $N=Z$ line. This observation may be a result of the strong spin-aligned $np$ pairing interaction which favors a three-quasi-particle-like configuration as the coupling between an aligned $np$ pair and an odd particle. Such effects of strong isoscalar pairing correlations are unique to odd-$A$ $N\sim Z$ nuclei like $^{87}$Tc.

\paragraph{Acknowledgement.}

This work was supported by the Swedish Research Council under Grant No. 621-2014-5558 and 2019-04880, the EU 7th Framework Programme, Integrating Activities Transnational Access, project No.262010 ENSAR, the UK STFC under grants ST/L005727/1 and ST/P003885/1, the Polish National Science Centre, grant no. 2017/25/B/ST2/01569, COPIN-INFN, COPIN-IN2P3 and COPIGAL projects, the National Research, Development and Innovation Fund of Hungary (Project no. K128947), the European Regional Development Fund (Contract No. GINOP-2.3.3-15-2016-00034), the Hungarian National Research, Development and Innovation Office - NKFIH, contract number PD124717, the Ministerio de Ciencia e Innovaci\'on and Generalitat Valenciana, Spanish, under the Grants SEV-2014-0398, FPA2017-84756-C4, PROMETEO/2019/005 and by the EU FEDER funds. X. L. gratefully acknowledges support from the China Scholarship Council, grant No. 201700260183 for his stay in Sweden. We thank the GANIL staff for excellent technical support and operation.

\bibliography{Ru}

\end{document}